\documentclass[aps,prd,showpacs,notitlepage,nofootinbib,superscriptaddress,floatfix,showkeys,twocolumn]{revtex4-1}
\usepackage{blindtext}
\usepackage{hyperref}
\usepackage{amsmath,amssymb}
\usepackage{float}
\usepackage{microtype}
\usepackage{balance}
\usepackage{comment}

\usepackage{graphicx}
\usepackage{bm}
\usepackage{latexsym}
\usepackage{epsfig}
\usepackage{psfrag}
\usepackage{xcolor}
\usepackage[dvipsnames]{xcolor}
\usepackage{subfigure}
\usepackage{graphicx}

\usepackage{amssymb}
\usepackage{amsmath}
\usepackage{bm}
\usepackage{latexsym}
\usepackage{epsfig}
\usepackage{psfrag}
\usepackage[normalem]{ulem}
\usepackage{textcomp}
\usepackage{color}
\usepackage{pstricks}
\usepackage[utf8]{inputenc}
\usepackage{physics}
\def\Mpl{M_{_{\mathrm{Pl}}}}

\def\d{\mathrm{d}}
\def\e{{\mathrm{e}}}
\def\nn{\nonumber}

\def\cs{c_{_\mathrm{S}}}

\def\vx{\bm{x}}
\def\vk{\bm{k}}
\def\ka{{k}_1}
\def\kb{{k}_2}
\def\kc{{k}_3}

\def\vka{\bm{k}_1}
\def\vkb{\bm{k}_2}

\def\vkd{\bm{k}_4}

\def\cL{\mathcal{L}}

\def\cO{\mathcal{O}}

\def\ea{\eta_1}
\def\eb{\eta_2}

\def\cs{c_{_\mathrm{S}}}

\begin{document}
\title{Massive Exchange and the Sign of the Equilateral Bispectrum}

\author{Diptimoy Ghosh }
\email{diptimoy.ghosh@iiserpune.ac.in}
\affiliation{Indian Institute of Science Education and Research, Pune}
\author{Suvashis Maity}
\email{saamaity@gmail.com}
\affiliation{Indian Institute of Science Education and Research, Pune}
\author{Farman Ullah }
\email{farman.ullah@icts.res.in}
\affiliation{International Centre for Theoretical Sciences-TIFR, Bengaluru}

% \date{}

\begin{abstract}

We study the inflationary bispectrum generated by the tree-level exchange of a massive hidden-sector scalar during inflation. 
When the interaction between the inflaton and the hidden sector arises only from the leading boost-breaking operator of the Effective Field Theory (EFT) of inflation, the equilateral bispectrum for principal-series scalar exchange is known to be universally negative, independent of the sign of the coupling. 
We revisit this result within the full EFT operator basis. Using bootstrap methods, we construct the de Sitter-invariant seed four-point function and obtain the inflationary bispectrum via weight-shifting operators and a soft-limit procedure. 
While the equilateral bispectrum remains strictly negative when only the leading interaction is present, additional operators generate independent cubic structures whose contributions compete in the equilateral configuration. As a result, the sign of the bispectrum is no longer universal. We derive a critical ratio of interaction coefficients that separates regions of positive and negative equilateral bispectrum.
We further study the effects of reduced sound speed $c_s<1$ and the exchange of multiple particles. In both cases the critical ratio is modified, and for multi-particle exchange a positive equilateral bispectrum can arise even when the higher-order operator is subdominant. 
Our results show that the negativity of the equilateral bispectrum from massive exchange is not generic, but reflects a restricted operator structure in the EFT of inflation.

\end{abstract}

\maketitle
% \tableofcontents
%%%%%%%%%%%%%%%%%%%%%%%%%%%%%%%%%%%%%%%%%%%%%%%%%%%%%%%%%%%%
%%%%%%%%%%%%%%%%%%%%%%%%%%%%%%%%%%%%%%%%%%%%%%%%%%%%%%%%%%%%

\section{Introduction}

\noindent Cosmic inflation \cite{Starobinsky:1980te,Guth:1980zm,Linde:1981mu,Starobinsky:1979ty,Mukhanov:1981xt,Martin:2003bt,Baumann:2009ds,Sriramkumar:2009kg} provides a compelling framework for explaining the observed large-scale homogeneity, isotropy, and flatness of the Universe, while simultaneously generating primordial fluctuations that seed the Cosmic Microwave Background (CMB) anisotropies and large-scale structure (LSS)~\cite{Martin:2003bt,Bassett:2005xm,Sriramkumar:2009kg,Baumann:2008bn}. 
Current CMB observations tightly constrain the scalar two-point function to be nearly scale invariant, with amplitude $\sim 2.09\times10^{-9}$ and spectral index $\sim 0.96$~\cite{Planck:2018vyg,Planck:2018jri,BICEP:2021xfz}. 
These observations constrain inflationary dynamics to some extent, however, a wide class of inflationary models can produce nearly identical power spectra. Therefore, one needs to analyse higher-point functions (non-Gaussianities)~\cite{Maldacena:2002vr,
Seery:2005wm,Chen:2006nt,Langlois:2008qf,Cheung:2007sv,Holman:2007na,Chen:2008wn,Shandera:2008ai,Chen:2009bc,Gao:2009at,Chen:2010xka,Senatore:2010jy,Bartolo:2010bj,Ghosh:2022cny,Ghosh:2023agt,Bhowmick:2024kld,Ghosh:2024aqd,Ghosh:2025pxn,Bhowmick:2025mxh,Ansari:2025nng}. 
Non-Gaussianities provide a crucial diagnostic for distinguishing between models. For instance, a large local non-Gaussianity is a smoking gun for the presence of additional fields during inflation.

% \vspace{0.2cm}

% \noindent
% {\bf Hidden sectors and inflationary collider physics.}

While the simplest realization of inflation involves a single scalar field minimally coupled to gravity, ultraviolet completions of inflation---particularly those with extra spacial dimensions---generically predict the existence of additional sectors~\cite{Baumann:2014nda,Chen:2010xka,Baumann:2011nk,Assassi:2012zq,Noumi:2012vr,Arkani-Hamed:2015bza,Maldacena:2002vr,Arkani-Hamed:2018kmz,Pimentel:2025rds,Jiang:2025mlm,Banks:1981nn,Georgi:2007ek,Georgi:2007si,Georgi:2008pq,Georgi:2009xq,Grinstein:2008qk,Stephanov:2007ry,Cespedes:2012hu,Wang:2025qww,Chakraborty:2023qbp,Craig:2024qgy}. 
Upon compactification from higher dimensions, extra gauge fields, moduli, heavy scalars, and other degrees of freedom naturally arise. 
If some of these fields have masses comparable to the Hubble scale during inflation, their presence can be indirectly probed through higher-point correlation functions~\cite{Alishahiha:2004eh,Babich:2004gb,Bartolo:2004if,Creminelli:2004yq,Seery:2005wm,Tolley:2009fg,Baumann:2011su,Achucarro:2010da,Flauger:2016idt}. 
Such sectors are commonly referred to as hidden sectors, as they interact only weakly with the inflaton, often gravitationally or through suppressed couplings.

The effects of these additional degrees of freedom can be systematically analyzed using the Effective Field Theory (EFT) of Inflation~\cite{Cheung:2007st,Weinberg:2008hq,Baumann:2009ds,Lee:2016vti,Weinberg:2009bg,Bartolo:2010bj,Senatore:2010wk,Creminelli:2010qf,Fasiello:2011fj,Jimenez:2011nn,Shiu:2011qw,LopezNacir:2011kk,OConnell:2011boq,Bloomfield:2011np,Achucarro:2012sm,Burgess:2012dz,Craig:2024qgy}. 
To leading order in derivatives, the EFT action in the unitary gauge takes the form
\begin{align}
    S &=\int \d^4x\sqrt{-g}\Big[\frac{1}{2}\Mpl^2 R
    +\Mpl^2\dot{H}g^{00} -\Mpl^2(3H^2+\dot{H}) \nn\\
    &\qquad
    +\sum_{n=2}^\infty \frac{M_n^4}{n!}(g^{00}+1)^n+...\Big],
    \label{eq:full-action}
\end{align}
where $g^{00}$ is the time-time component of the metric, $H$ is the Hubble parameter and $R$ is the four dimensional Ricci scalar.
In the pure EFT of inflation there is no independent linear operator in 
$\delta g^{00}\equiv g^{00}+1$; its coefficient is fixed by the background equations, 
and the first free deformations begin at quadratic order.  
Restoring time diffeomorphisms via the Goldstone mode $\pi$ under 
$t\to t+\pi(t,\vx)$ yields
\begin{align}
    g^{00} \to -1-2\dot\pi + (\partial_\mu\pi)^2 ,
\end{align}
where we have written the result in the decoupling limit. The fluctuations of the inflaton are related to the Goldstone by 
$\pi=\delta\phi/\dot\phi$, and the curvature perturbation, $\zeta$ is defined as
\begin{align}
    \zeta = -H\pi + O(\pi^2).
\end{align}
Note that, $\zeta$ is the quantity, whose statistics determine the anisotropies observed in the CMB.

% \vspace{0.2cm}

% \noindent
% {\bf Universality of the equilateral sign in weakly broken de Sitter symmetry.}

Before turning to hidden-sector exchange, it is useful to recall a simpler case. 
In a shift-symmetric EFT, the leading higher-derivative interaction of the inflaton can be written as~\cite{Creminelli:2003iq}
\begin{align}
\cL=\sqrt{-g}\,\frac{\alpha}{\tilde\Lambda^4}(\partial\phi)^4.
\end{align}
Flat-space positivity arguments imply $\alpha>0$~\cite{Pham:1985cr,Adams:2006sv,deRham:2025mjh}. 
Taking one external mode to be soft generates an effective cubic interaction that produces a bispectrum, which is negative in the equilateral configuration~\cite{Creminelli:2003iq}. 
In this case, the sign of the bispectrum follows directly from positivity and kinematics.

A closely related phenomenon occurs for the tree-level exchange of a massive scalar belonging to the principal series ($m^2>9/4H^2$). 
It has been shown~\cite{deRham:2025mjh} that, in the regime where de Sitter symmetry is only weakly broken, the equilateral bispectrum arising from massive exchange is negative, independently of the sign of the interaction coefficient. 
In this regime, the sign is controlled entirely by kinematical properties of the de Sitter-invariant seed correlator.

% \vspace{0.2cm}

% \noindent
% {\bf Relation to previous work.}
Recent bootstrap- and positivity-based approaches have developed a general framework to constrain inflationary correlators sourced by hidden sectors. The have emphasized the role of de Sitter symmetry, analyticity and unitarity of the hidden-sector two-point function, and the utility of de Sitter-invariant ``seed'' correlators combined with differential/weight-shifting maps to inflationary observables \cite{Arkani-Hamed:2015bza,Arkani-Hamed:2018kmz,Jazayeri:2022kjy,Wang:2025qww,Baumann:2020dch,Karateev:2017jgd,Hogervorst:2021uvp,Baumann:2019oyu}. 
Within this framework, the universal negativity of the equilateral bispectrum for the exchange of scalar particles in the principal series arises naturally when the cubic structure is effectively restricted to that generated by the leading order coupling of the clock field to the hidden sector.

\begin{figure}
    \centering
    \includegraphics[width=\linewidth]{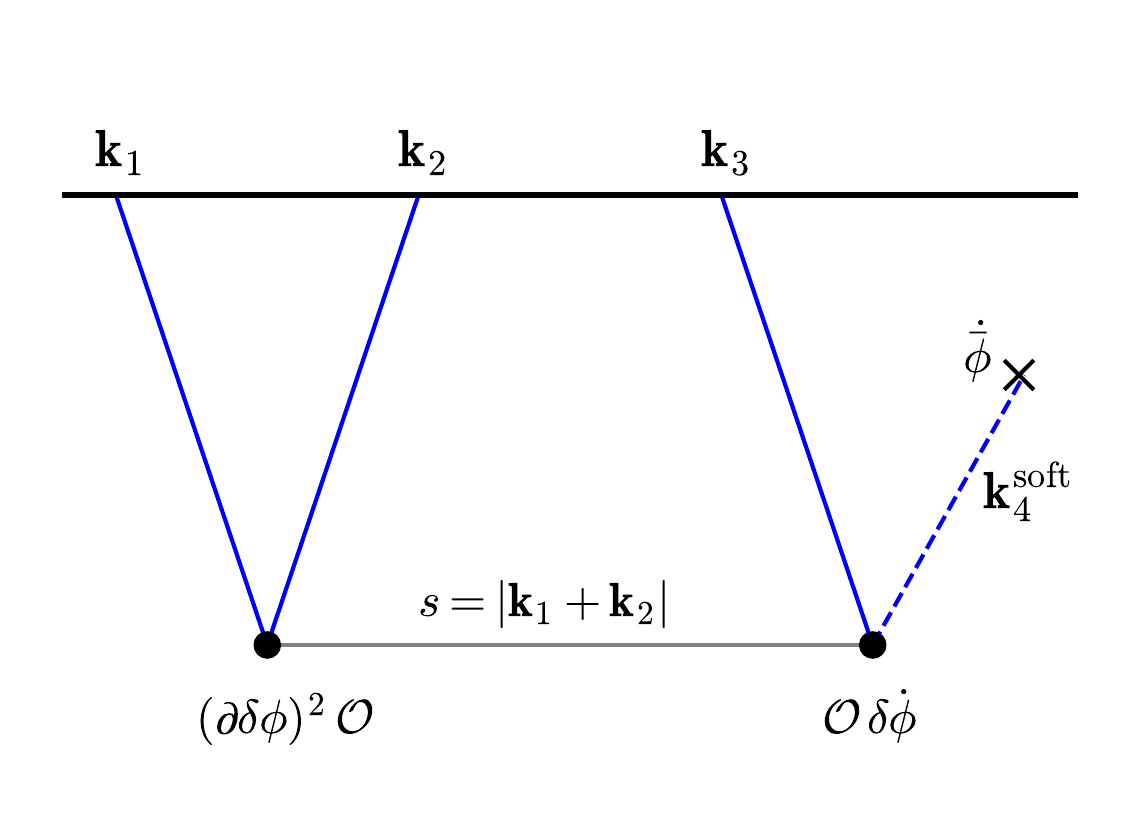}
    \caption{
    The four-point function with the exchange of a tree-level massive hidden-sector field is illustrated here. The hidden sector is denoted in a gray line, whereas the external legs corresponding to the inflaton are given in blue lines. 
    The inflationary bispectrum is obtained by evaluating one external leg in the soft limit, 
    corresponding to the background mode $k_4\to0$. 
    This soft-limit procedure maps the de Sitter-invariant seed correlator 
    to the inflationary three-point function.
    }
    \label{fig:soft-mode}
\end{figure}

At the same time, it has been stressed that once one works in the general EFT of inflation operator basis, additional unitary-gauge building blocks such as $(\delta g^{00})^2 \cO$ (or equivalently $(g^{00}+1)^2\cO$) are allowed and need not be parametrically suppressed relative to $\delta g^{00}\cO$ \cite{deRham:2025mjh}. 
The general implication is that sign universality is not guaranteed across the full EFT parameter space. 
Our goal in this paper is to make this loss of universality completely explicit and quantitatively sharp for the exchange process of massive principal-series scalar(s) (see Fig.~\ref{fig:soft-mode}).

\begin{comment}
\vspace{0.2cm}

\noindent
{\bf Question addressed in this work.}\textcolor{red}{looks redundant}

The universality of this negative sign relies crucially on the assumption that de Sitter symmetry is only weakly broken. 
The full EFT of inflation, however, contains an infinite tower of operators of the form $(g^{00}+1)^n$, which enlarge the space of allowed boost-breaking deformations at fixed derivative order. 
Once such operators are included, distinct cubic structures for the Goldstone field appear, with independent relative weights between time- and space-derivative interactions.
This raises a natural question:

\begin{quote}
\emph{Does the universal negativity of the equilateral bispectrum from massive exchange persist once the full set of boost-breaking operators allowed by the EFT is included?}
\end{quote}

In this work we show that the answer is negative. 
When higher-order operators are included, the equilateral bispectrum receives competing contributions from distinct cubic interactions. 
As a result, its sign is no longer universal. 
Instead, we derive an explicit bound on the ratio of interaction coefficients that determines when the equilateral bispectrum becomes positive.
\end{comment}
% \vspace{0.2cm}

% \noindent
% {\bf Setup and main result.}

We focus on the case where the hidden sector consists of a massive scalar $\sigma$ exchanged at tree level (see Fig.~\ref{fig:soft-mode}). Since it is not straightforward to proceed via direct in-in calculations, we use bootstrap methods and symmetry constraints. We first consider the de Sitter seed four-point function~\cite{Arkani-Hamed:2018kmz}. 
The inflationary bispectrum is then obtained by applying appropriate weight-shifting operators and taking one external leg soft~\cite{Arkani-Hamed:2018kmz,Baumann:2019oyu}.

%%%%%%%%%%%%%%%%%%%%%%%%%%%%%%%%%%%%%%%%%%%%%%%%%%
%%%%%%%%%%%%%%%%%%%%%%%%%%%%%%%%%%%%%%%%%%%%%%%%%%

We show that:
\begin{itemize}
    \item When the cubic structure is restricted to arise from the linear operator $(g^{00}+1)\sigma$, the equilateral bispectrum is strictly negative for all real values of the mass parameter $\mu$ in the principal series.
    \item Once additional operators allowed by the EFT are included, the bispectrum receives multiple contributions with distinct derivative structures.
    \item The sign of the equilateral configuration is then controlled by the competition between these contributions, leading to a calculable bound on the ratio of interaction coefficients.
    \item The critical ratio depends on the sound speed $\cs$ of the Goldstone mode and is significantly modified when $\cs<1$. In addition, when multiple hidden-sector particles are exchanged, their collective contributions can further modify the sign condition, allowing a positive equilateral bispectrum even in regions where the higher-order operator coefficient satisfies $c_2<c_1$ (see, Eq.~\eqref{eq:lint1}).
\end{itemize}

Thus, the negativity of the equilateral bispectrum is not a generic prediction of massive exchange, but rather a consequence of restricted symmetry-breaking structure. 
These results suggest that the sign of the equilateral bispectrum itself can serve as a diagnostic of the symmetry-breaking structure of the EFT of inflation and of the spectrum of heavy particles coupled to the inflaton.

\vspace{0.2cm}

The rest of the paper is organized as follows.
In Sec.~\ref{sec:seed-4pt}, we review the seed four-point function for massive exchange in the principal series. We also discuss about the weight shifting operators that translate the case of conformally coupled scalars to massless scalars.
In Sec.~\ref{sec:bispec-massive}, we compute the inflationary bispectrum and determine the conditions under which its equilateral limit changes sign. We discuss the effects of different speeds of sound, $\cs$, and also the case of the exchange of multiple particles.
Finally, in Sec.~\ref{sec:conclusion}, we conclude with a brief summary and future outlook.

\vspace{0.2cm}

\noindent
{\bf Notational conventions:}
We work in natural units $\hbar=c=1$ and set $\Mpl=(8\pi G)^{-1/2}$.
We adopt the metric signature $(-,+,+,+)$ and assume a spatially flat FLRW background,
\begin{align}
    \d s^2=a^2 (-\d\eta^2+\d\vx^2),
\end{align}
with $H=\dot a/a$.
An overdot denotes differentiation with respect to cosmic time $t$, and an overprime with respect to conformal time $\eta$.
We use the notation $k_{ij}=k_i+k_j$.
%%%%%%%%%%%%%%%%%%%%%%%%%%%%%%%%%%%%%%%%%%%%%%%%%%%%%%%%%%%%
%%%%%%%%%%%%%%%%%%%%%%%%%%%%%%%%%%%%%%%%%%%%%%%%%%%%%%%%%%%%
\section{Four-point seed function}
\label{sec:seed-4pt}

As discussed in the previous section, our goal is to determine the inflationary bispectrum generated by the exchange of a massive hidden-sector field. 
Rather than computing the three-point function directly, it is convenient to first evaluate the corresponding four-point function in de Sitter space. 
The inflationary bispectrum is then obtained by taking one of the external legs of the trispectrum to be soft (see Fig.~\ref{fig:soft-mode}). 

%%%%%%%%%%%%%%%%%%%%%%%%%%%%%%%%%%%%%%%%%%%%%%%%%%%%%%%%%%%%%%%%%%%%%%%%
% \subsection*{Seed correlator and weight shifting}
\subsection*{The four-point correlation}
%%%%%%%%%%%%%%%%%%%%%%%%%%%%%%%%%%%%%%%%%%%%%%%%%%%%%%%%%%%%%%%%%%%%%%%%

In the inflationary trispectrum, the external fields are massless scalars $\phi$. 
However, it is technically simpler to first compute the four-point function with conformally coupled scalars $\varphi$ as external modes. 
The correlator for massless scalar perturbations can subsequently be obtained by applying appropriate weight-shifting operators~\cite{Arkani-Hamed:2018kmz,Baumann:2019oyu}.

The four-point function in the in-in formalism is defined as 
\begin{align}
    \langle \varphi_1\varphi_2\varphi_3\varphi_4 \rangle
    =
    (2\pi)^3\,\delta^3\!\left(\sum_i^4 \vk_i\right) 
    \sum_{a,b=\pm}I_{ab},
\end{align}
where
\begin{align}
    I_{\pm\pm}
    &=
    (\pm ig)(\pm ig)\kappa 
    \int\!\!\int
    \frac{\d\ea}{\ea^2}
    \frac{\d\eb}{\eb^2}
    \nn\\&\quad\times
    \e^{\pm i k_{12}\ea}
    G_{\pm\pm}(s,\ea,\eb)
    \e^{\pm i k_{34}\eb}.
    \label{eq:timeintegmassive}
\end{align}
The delta function in the above expression ensures the conservation of the external momentum. The factor $\pm ig$ is the multiplicative term corresponding to each vertex, where we shall take $g=1$.
Further, $\kappa=8\pi^3 (|\dot{\bar\phi}|/\Lambda^2) \mathcal{P}_\zeta^{3/2}$ involves various quantities with $\mathcal{P}_\zeta\sim 2.1\times10^{-9}$ being the scalar power spectrum.
The propagators $G$'s are two-point correlators of the hidden-sector field, given by 
\begin{align}
    G_{++}(k,\ea,\eb)
    &= f(k,\ea)f^\ast(k,\eb)\theta(\ea-\eb)
    +{\ea\rightleftharpoons\eb}, \\
    G_{+-}(k,\ea,\eb)
    &= f^\ast(k,\ea)f(k,\eb), \\
    G_{--}(k,\ea,\eb)
    &= G_{++}^\ast(k,\ea,\eb), \\
    G_{-+}(k,\ea,\eb)
    &= G_{+-}^\ast(k,\ea,\eb),
\end{align}
where $f$ denotes the appropriate mode function.
The mode functions for a conformally coupled scalar and the Goldstone are
\begin{align}
    \varphi^\pm(\cs k,\eta)
    &= -\frac{H}{\sqrt{2\cs k}}\eta\,\e^{\mp i\cs k\eta}, \\
    \pi^\pm(\cs k,\eta)
    &= -\frac{H}{\sqrt{2\cs^3k^3}}
    (1\pm i\cs k\eta)\,\e^{\mp i \cs k\eta}.
\end{align}
From these expressions, it follows that
\begin{align}
    \pi_k=(\cs k\eta)^{-1}\left[1-k\partial_k\right]\varphi_k,
\end{align}
which defines the weight-shifting operator mapping the conformally coupled particle to the massless particle.
In what follows, we now focus on the seed four-point function with conformally coupled external fields.

%%%%%%%%%%%%%%%%%%%%%%%%%%%%%%%%%%%%%%%%%%%%%%%%%%%%%%%%%%%%%%%%%%%%%%%%
\subsection*{Massive hidden-sector field}
%%%%%%%%%%%%%%%%%%%%%%%%%%%%%%%%%%%%%%%%%%%%%%%%%%%%%%%%%%%%%%%%%%%%%%%%

Let us consider a hidden-sector scalar of mass $m$. The corresponding mass function is defined as 
\begin{align}
    \mu=\sqrt{\frac{m^2}{H^2}-\frac{9}{4}},
    \qquad
    \Delta_\sigma=\frac{3}{2}+ i\mu,
\end{align}
where $\Delta_\sigma$ is the asociated scaling dimension.
For $m\geq 3H/2$, $\mu$ is real and the field belongs to the principal series; 
for $m<3H/2$, $\mu$ becomes imaginary and defines the complementary series. 
We focus on the principal-series regime, \textit{corresponding to a gapped hidden sector} whose excitations decay outside the horizon\footnote{To consider a gapped hidden sector, one can in principle start with any nonzero mass for the hidden sector particles, $m>0$. However, our results also depend on the mass gap. Our aim is to focus on scenarios in which the hidden sector particles decay outside the Hubble radius, leaving the massless scalar field as the only observable field in the CMB. Choosing the mass gap as the starting point of the principal series is therefore a well-motivated choice, as it ensures the decay of the hidden sector particles.}.
The mode functions for a principal-series scalar are given by
\begin{align}
    \sigma^+(k,\eta)
    &= \frac{\sqrt{\pi}H}{2}
    \e^{-\pi\mu/2+i\pi/4}
    (-\eta)^{3/2}
    H_{i\mu}^{(1)}(-k\eta), \\
    \sigma^-(k,\eta)
    &= \frac{\sqrt{\pi}H}{2}
    \e^{\pi\mu/2-i\pi/4}
    (-\eta)^{3/2}
    H_{i\mu}^{(2)}(-k\eta),
\end{align}
with $H_\nu^{(1)}$, $H_\nu^{(2)}$ being the Hankel functions of first and second kind.
In the next section, we shall discuss the method of computing the seed four-point function for a massive scalar exchange.

%%%%%%%%%%%%%%%%%%%%%%%%%%%%%%%%%%%%%%%%%%%%%%%%%%%%%%%%%%%%%%%%%%%%%%%%
\subsection*{Bootstrap differential equation}
%%%%%%%%%%%%%%%%%%%%%%%%%%%%%%%%%%%%%%%%%%%%%%%%%%%%%%%%%%%%%%%%%%%%%%%%

As already discussed, rather than evaluating the time integrals in Eq.~\eqref{eq:timeintegmassive} directly, it is convenient to determine the four-point function by exploiting the differential equations implied by de Sitter symmetry.
For $s$-channel exchange, with $s=\vert \vka+\vkb\vert$, we introduce dimensionless variables
\begin{align}
   u=\frac{s}{\cs k_{12}}, 
   \qquad 
   v=\frac{s}{\cs k_{34}}.
\end{align}
These variables lie on a disc of radius $c^{-1}_s$ in the $u$-$v$ plane.
De Sitter invariance implies that the four-point function $\hat F(u,v)$ satisfies the following differential equations
\begin{align}
    \left(\Delta_u+\mu^2+\frac{1}{4}\right)\hat{F}(u,v)
    &= g^2\frac{uv}{u+v}, \nn \\
    \left(\Delta_v+\mu^2+\frac{1}{4}\right)\hat{F}(u,v)
    &= g^2\frac{uv}{u+v},
    \label{eq:diff-F}
\end{align}
where the operator
\begin{align}
    \Delta_u=u^2(1-u^2)\partial_u^2-2u^3\partial_u.
    \label{eq:deltamu1}
\end{align}
The source term on the right-hand side of Eq.~\eqref{eq:diff-F} encodes the local cubic interaction.
Solutions to the homogeneous equation correspond to contact terms, 
while the inhomogeneous solution captures the non-local exchange contribution. Further, from Eq.~\eqref{eq:diff-F}, it is also clear that the solution of $\hat{F}(u,v)$ will be symmetric in $u$ and $v$.

%%%%%%%%%%%%%%%%%%%%%%%%%%%%%%%%%%%%%%%%%%%%%%%%%%%%%%%%%%%%%%%%%%%%%%%%
\subsection*{Structure of the solution}
%%%%%%%%%%%%%%%%%%%%%%%%%%%%%%%%%%%%%%%%%%%%%%%%%%%%%%%%%%%%%%%%%%%%%%%%

The solution of Eq.~\eqref{eq:diff-F} has been derived in detail in~\cite{Arkani-Hamed:2018kmz}. 
We write the solution here, which is as follows
\begin{align}
    \hat F(u,v)
    =
    \begin{cases}
        \displaystyle
        \sum_{m,n}c_{mn} u^{2m+1}\left(\frac{u}{v}\right)^n\\
        +
        \frac{\pi}{2\cosh(\pi\mu)}\hat{g}(u,v),
        & u\leq v, \\
        u\leftrightarrow v,
        & u\geq v,
    \end{cases}
\end{align}
where the first term corresponds to analytic contact contributions.
The second term $\hat g(u,v)$ contains the non-analytic structure characteristic of massive exchange.
Explicitly, $\hat{g} (u,v)$ is given by 
\begin{align}
    \hat{g} (u,v)
    &= \hat{F}_{+}(u)\hat{F}_{-}(v)
    -\hat{F}_{-}(u)\hat{F}_{+}(v) 
    \nn\\
    &\quad
    -\frac{\alpha_{-}}{\alpha_+}(\beta_0+1)
    \hat{F}_{+}(u)\hat{F}_{+}(v)
    \nn\\
    &\quad
    -\frac{\alpha_{+}}{\alpha_-}(\beta_0-1)
    \hat{F}_{-}(u)\hat{F}_{-}(v)
    \nn\\
    &\quad
    +\beta_0\left[
    \hat{F}_{+}(u)\hat{F}_{-}(v)
    +\hat{F}_{-}(u)\hat{F}_{+}(v)
    \right],
\end{align}
with
\begin{align}
\beta_0&=\frac{1}{i\sinh(\pi\mu)},\\
\alpha_{\pm} 
&= -\left(\frac{i}{2\mu}\right)^{\frac{1}{2}\pm i\mu}
\frac{\Gamma(1\pm i\mu)}
{\Gamma\left(\frac{1}{4}\pm i\frac{\mu}{2}\right)
\Gamma\left(\frac{3}{4}\pm i\frac{\mu}{2}\right)},\\
\hat{F}_{\pm}(u)
&=
\left(\frac{iu}{2\mu}\right)^{\frac{1}{2}\pm i\mu}
{}_2F_1\!\left(
\frac{1}{4}\pm i\frac{\mu}{2},
\frac{3}{4}\pm i\frac{\mu}{2};
1\pm i\mu; u^2
\right),
\end{align}
where ${}_2F_1$ is the hypergeometric function.

All non-analytic dependence on $u$ arises from this exchange term.
In particular, in the limit $u\to 0$ one finds
\begin{align}
\hat F(u,v)\sim u^{\frac{3}{2}\pm i\mu},
\end{align}
which is responsible for the oscillatory behavior characteristic of massive exchange in the squeezed limit of the bispectrum.
Thus the non-analytic part of the seed correlator is completely fixed by de Sitter symmetry and the mass parameter $\mu$, 
while analytic contact terms encode local ambiguities.

%%%%%%%%%%%%%%%%%%%%%%%%%%%%%%%%%%%%%%%%%%%%%%%%%%%%%%%%%%%%
%%%%%%%%%%%%%%%%%%%%%%%%%%%%%%%%%%%%%%%%%%%%%%%%%%%%%%%%%%%%

\section{Bispectrum for the exchange of a massive scalar}
\label{sec:bispec-massive}

\begin{figure*}
    \centering
    \includegraphics[width=0.45\linewidth]{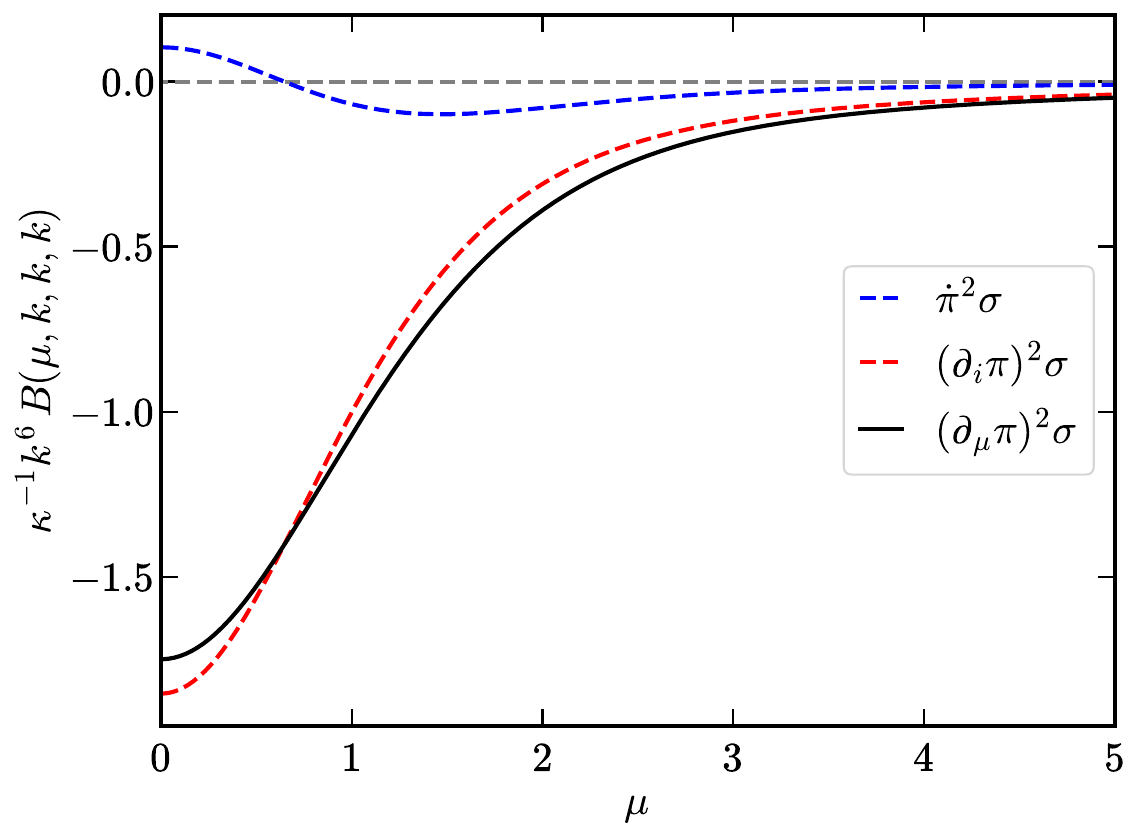}
    \includegraphics[width=0.45\linewidth]{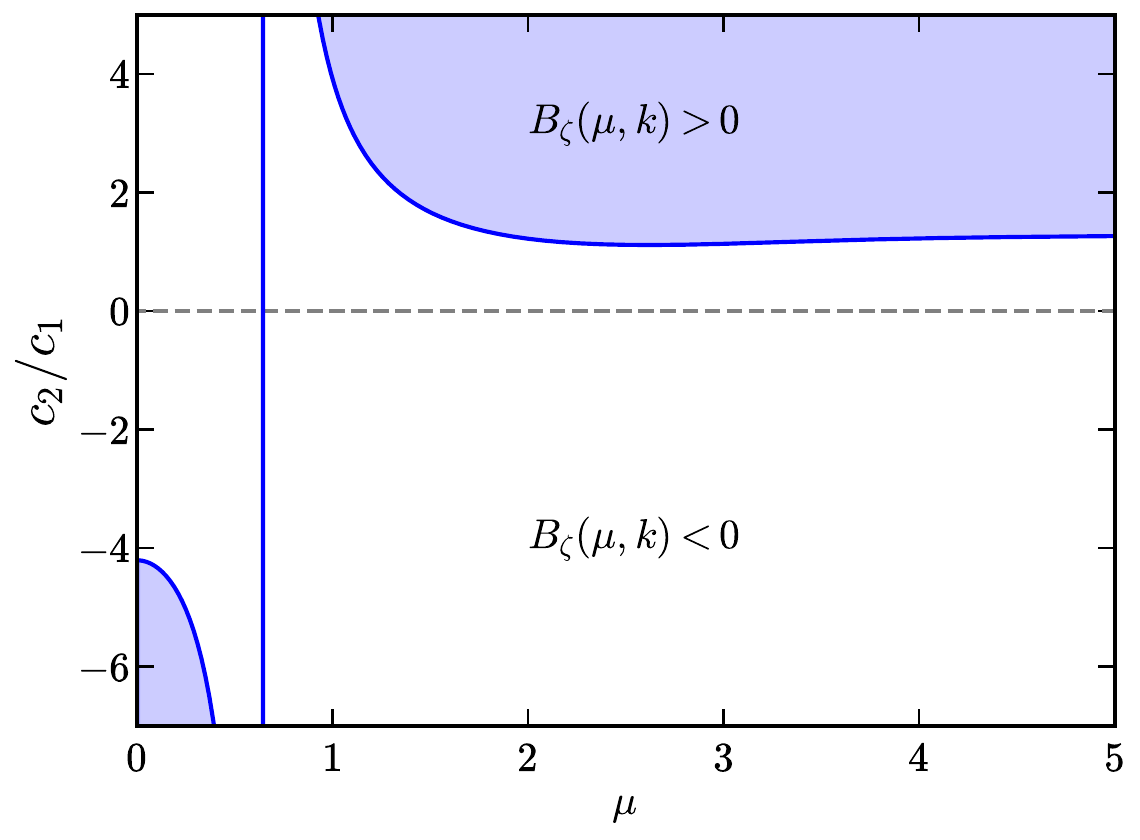}
    \caption{
    Purely kinematical part (without coefficients) of the bispectrum in equilateral limit 
    for the cubic structures $(\partial_\mu\pi)^2\sigma$ (in black), 
    $\dot\pi^2\sigma$ (in blue dashed), and $(\partial_i\pi)^2\sigma$ (in red dashed) are plotted on the left panel. The black line is obtained by adding the contributions from the blue and red lines. 
    On the right panel, we have plotted the critical ratio $c_2/c_1$, separating regions where the equilateral 
    bispectrum $B_\zeta(\mu,k)$ is positive or negative.
    }
    \label{fig:ratio-massive}
\end{figure*}

In this section, we shall discuss the solution of the seed four-point function, the structure of the interactions, and the conditions to obtain a positive bispectrum in the equilateral configuration.
%%%%%%%%%%%%%%%%%%%%%%%%%%%%%%%%%%%%%%%%%%%%%%%%%%%%%%%%%%%%%%%%%%%%%%%%
\subsection*{Cubic structures from the EFT of inflation}
%%%%%%%%%%%%%%%%%%%%%%%%%%%%%%%%%%%%%%%%%%%%%%%%%%%%%%%%%%%%%%%%%%%%%%%%

The interaction between the inflaton and a hidden-sector scalar $\sigma$ arises from the EFT of inflation operators
\begin{align}
    \cL_{\rm int}
    &= \sqrt{-g}\,c_1(g^{00}+1)\sigma
    +\sqrt{-g}\,c_2(g^{00}+1)^2\sigma+... \nn\\
    &= \sqrt{-g}\Big\{c_1[-2\dot\pi+(\partial_\mu\pi)^2]
    + c_2[-2\dot\pi+(\partial_\mu\pi)^2]^2\Big\}\sigma .
    \label{eq:lint1}
\end{align}
Expanding to quadratic order in $\pi$, we obtain
\begin{align}
    \cL_{\rm int}
    = \sqrt{-g}
    \Big[
    -2c_1\dot\pi
    -(c_1-4c_2)\dot\pi^2
    +c_1(\partial_i\pi)^2
    +\cO(\pi^3)
    \Big]\sigma .
    \label{eq:Lint1}
\end{align}
From the above equation, it is straightforward to see that there are two distinct cubic structures that contribute to the bispectrum:
\begin{itemize}
    \item A time-derivative interaction $\dot\pi^2\sigma$,
    \item A spatial-derivative interaction $(\partial_i\pi)^2\sigma$.
\end{itemize}
These operators carry different momentum dependence and therefore contribute differently to the equilateral configuration. 
Importantly, the operator quadratic  in $\delta g^{00}$ modifies the relative weight between the time- and space-derivative structures at fixed derivative order, thereby enlarging the space of allowed boost-breaking deformations within the EFT.

%%%%%%%%%%%%%%%%%%%%%%%%%%%%%%%%%%%%%%%%%%%%%%%%%%%%%%%%%%%%%%%%%%%%%%%%
\subsection*{From the seed correlator to the bispectrum}
%%%%%%%%%%%%%%%%%%%%%%%%%%%%%%%%%%%%%%%%%%%%%%%%%%%%%%%%%%%%%%%%%%%%%%%%

Since the external field in the observable correlator is the massless mode $\pi$, we apply the appropriate weight-shifting operator to the seed four-point function following the conversion derived in Sec.~\ref{sec:seed-4pt} (see~\cite{Arkani-Hamed:2018kmz,Baumann:2019oyu,Jazayeri:2022kjy} for a detailed derivation). 
The bispectrum is obtained by evaluating one external leg of the four-point function in the soft limit, say $\vkd\to0$, which corresponds to $v=1/\cs$. To start with, we first focus on the case $\cs=1$.
In this limit, the seed function reduces to~\cite{deRham:2025mjh}
\begin{align}
    \hat F(u,v=1)
    &= \left(\mu^2+\frac{1}{4}\right)^{-1}
    \frac{u}{1+u}
    \nn\\
    &\quad\times
    {}_3F_2\!\left(1,1,1;\frac{3}{2}+i\mu,\frac{3}{2}-i\mu;
    \frac{2u}{1+u}\right)
    \nn\\
    &\quad
    +\frac{\pi}{2\cosh(\pi\mu)}\hat g(u,1).
\end{align}
Now, we shall apply the weight shifting operator to the seed four-point function where the bispectrum decomposes into two contributions~\cite{Arkani-Hamed:2018kmz,Baumann:2019oyu,Jazayeri:2022kjy} as 
\begin{align}
    B_{\zeta}^{(1)} (\mu,\ka,\kb,\kc)
    &= \kappa 
    % (\ka\kb\kc)^2
    \frac{-\alpha_1}{\ka\kb\kc(\ka+\kb)^3}
    \nn\\&\times
    (2\partial_u+u\partial_u^2)\hat F,
    \\
    B_{\zeta}^{(2)} (\mu,\ka,\kb,\kc)
    &= \kappa 
    % (\ka\kb\kc)^2
    \frac{\alpha_2\cs\,\vka\cdot\vkb}{\ka^3\kb^3\kc^2}
    \nn\\
    &\times
    \left[1+u\partial_u+
    \frac{\cs^2\ka\kb}{\kc^2}
    u^3(2\partial_u+u\partial_u^2)\right]\hat F,
\end{align}
where $B_{\zeta}^{(1)} (\mu,\ka,\kb,\kc)$ and $B_{\zeta}^{(2)} (\mu,\ka,\kb,\kc)$ correspond to the contribution from the interaction $\dot\pi^2\sigma$ and $(\partial_i\pi)^2\sigma$, respectively.
 The function $\hat F$ is the seed four-point function derived before, on which the weight shifting operators act.
In general, one further defines a dimensionless shape function which is given by $\equiv(\ka\kb\kc)^2 B_{\zeta}(\mu,\ka,\kb,\kc)$, that we shall also plot in this paper.
The coefficients are given by
\begin{align}
    \alpha_1 = c_1(c_1 - 4c_2), 
    \qquad
    \alpha_2 = c_1^2 .
\end{align}
These coefficients are important in our context, as we shall see in that the relative strengths of $c_1$ and $c_2$ determine the sign of the equilateral bispectrum.
In the following section, we shall discuss the role of the strength of the coefficients on the results.

%%%%%%%%%%%%%%%%%%%%%%%%%%%%%%%%%%%%%%%%%%%%%%%%%%%%
\begin{figure*}[t!]
    \centering
    \includegraphics[width=0.49\linewidth]{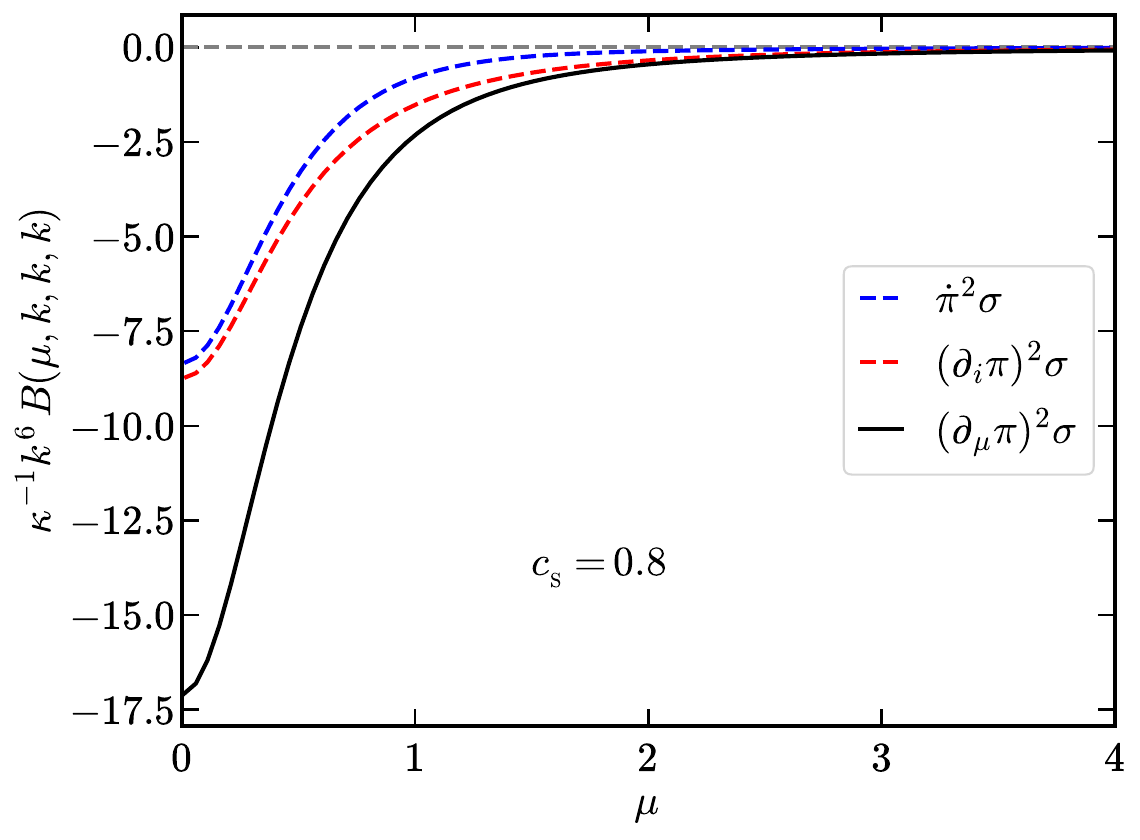}
    \includegraphics[width=0.49\linewidth]{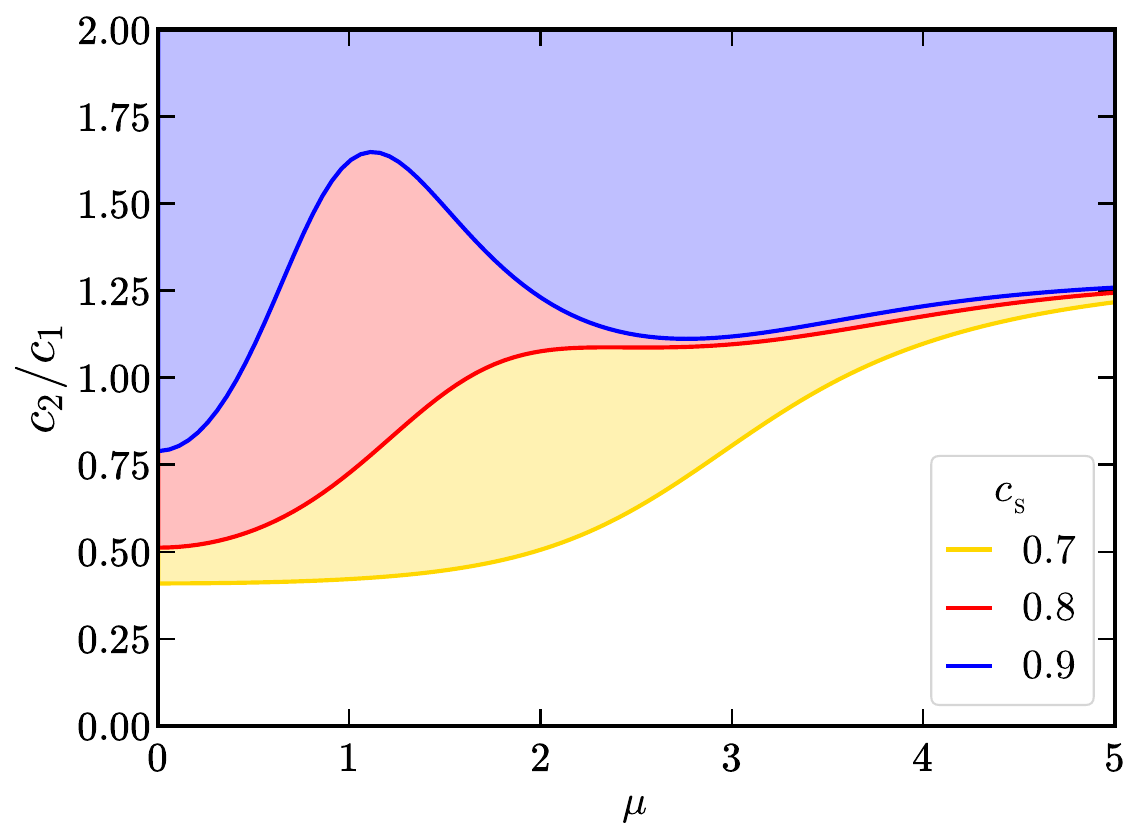}
    \caption{Purely kinematical part (without coefficients) of the bispectrum in equilateral limit 
    for the cubic structures $(\partial_\mu\pi)^2\sigma$ (in black), 
    $\dot\pi^2\sigma$ (in blue dashed), and $(\partial_i\pi)^2\sigma$ (in red dashed) are plotted on the left panel for $\cs=0.8$. 
    The critical ratio of $c_2/c_1$ separating regions where the equilateral 
    bispectrum $B_\zeta(\mu,k)$ is positive or negative are plotted in the right panel for the cases of $\cs=0.9$, $0.8$, $0.7$ in blue, red, and yellow, respectively. }
    \label{fig:csl1}
\end{figure*}
%%%%%%%%%%%%%%%%%%%%%%%%%%%%%%%%%%%%%%%%%%%%%%%%%%%%

%%%%%%%%%%%%%%%%%%%%%%%%%%%%%%%%%%%%%%%%%%%%%%%%%%%%%%%%%%%%%%%%%%%%%%%%
\subsection*{Squeezed limit and collider signal}
%%%%%%%%%%%%%%%%%%%%%%%%%%%%%%%%%%%%%%%%%%%%%%%%%%%%%%%%%%%%%%%%%%%%%%%%

In the squeezed configuration $\ka\sim\kb=k\gg\kc$, the bispectrum takes the form
\begin{align}
    B_{\zeta}
    =
    -\frac{\kappa\pi\mu}{k^6}
    \left(\frac{\kc}{k}\right)^{-3/2+i\mu}
    \mathcal{C}(\mu)
    + \text{c.c.},
\end{align}
where $\mathcal{C}(\mu)$ collects the $\mu$-dependent prefactors.
The oscillatory behavior $\sim (k_c/k)^{\pm i\mu}$ is the characteristic collider signal of a massive particle in de Sitter space.
It is well known that the amplitude is exponentially suppressed for large $\mu$~\cite{Arkani-Hamed:2018kmz,deRham:2025mjh}.

%%%%%%%%%%%%%%%%%%%%%%%%%%%%%%%%%%%%%%%%%%%%%%
\begin{figure*}
    \centering
    \includegraphics[width=0.49\linewidth]{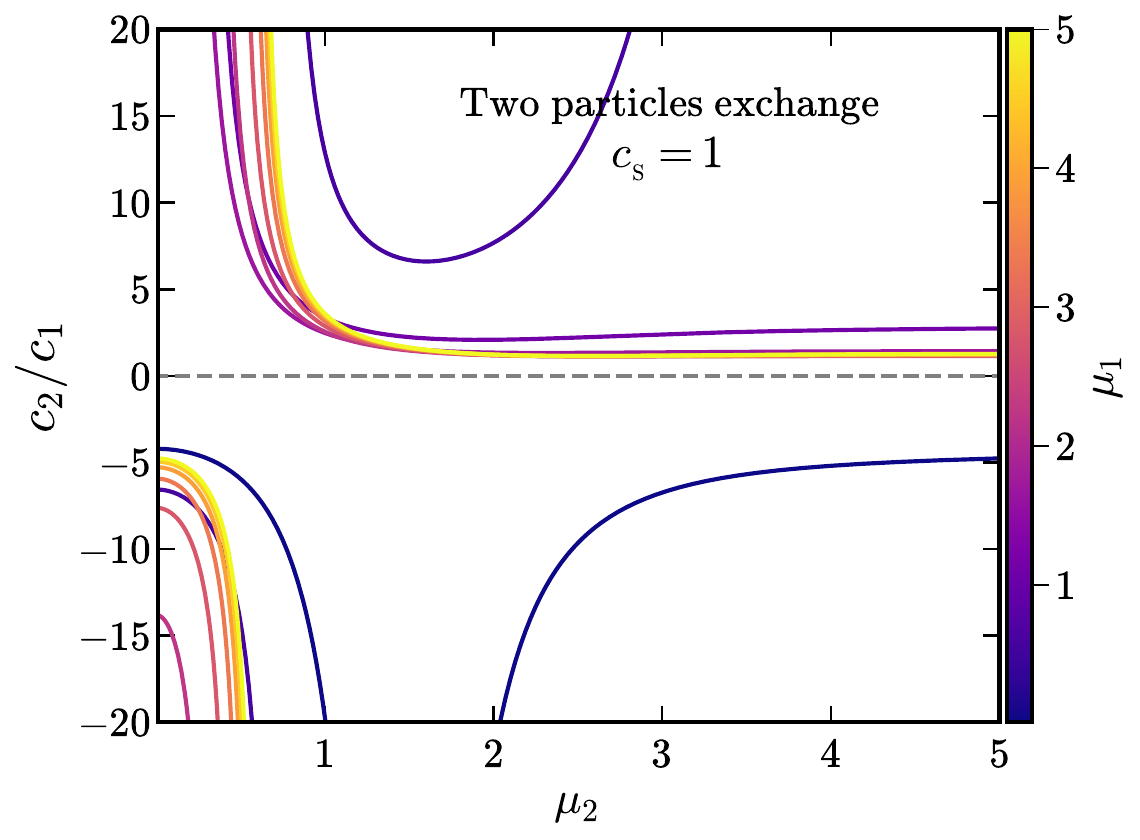}
    \includegraphics[width=0.49\linewidth]{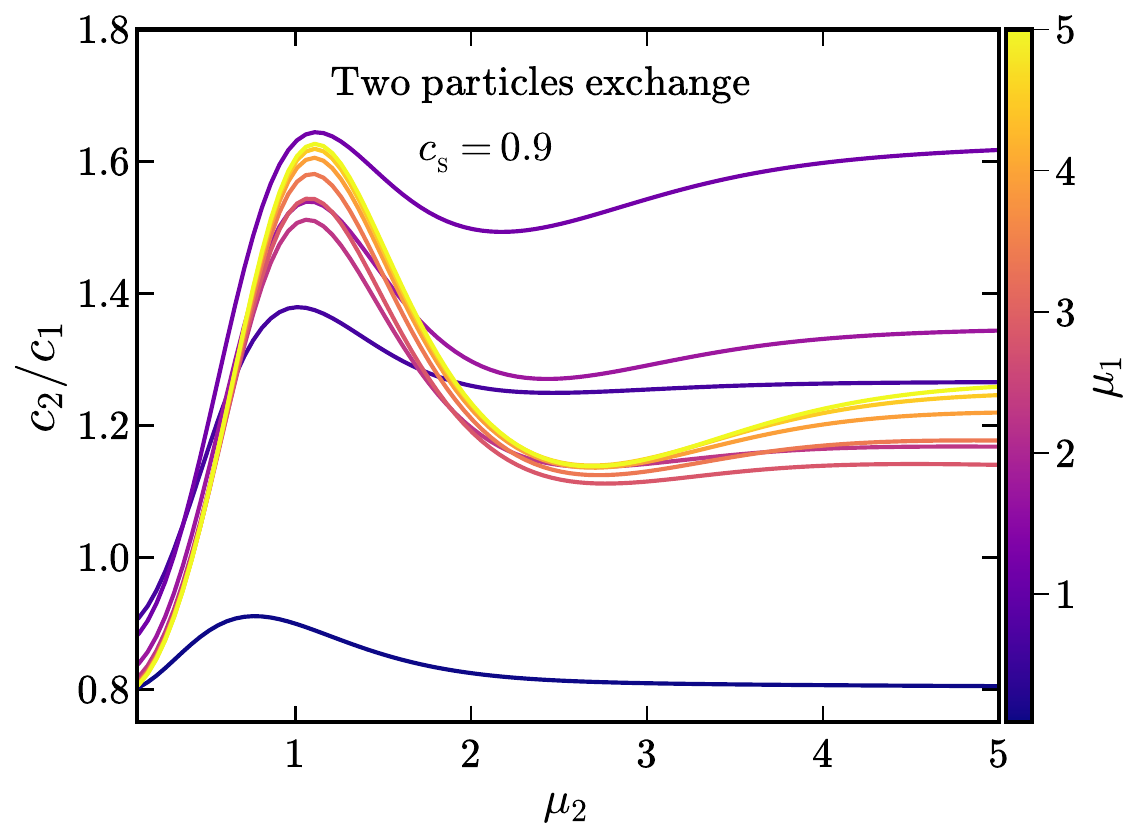}
    \includegraphics[width=0.49\linewidth]{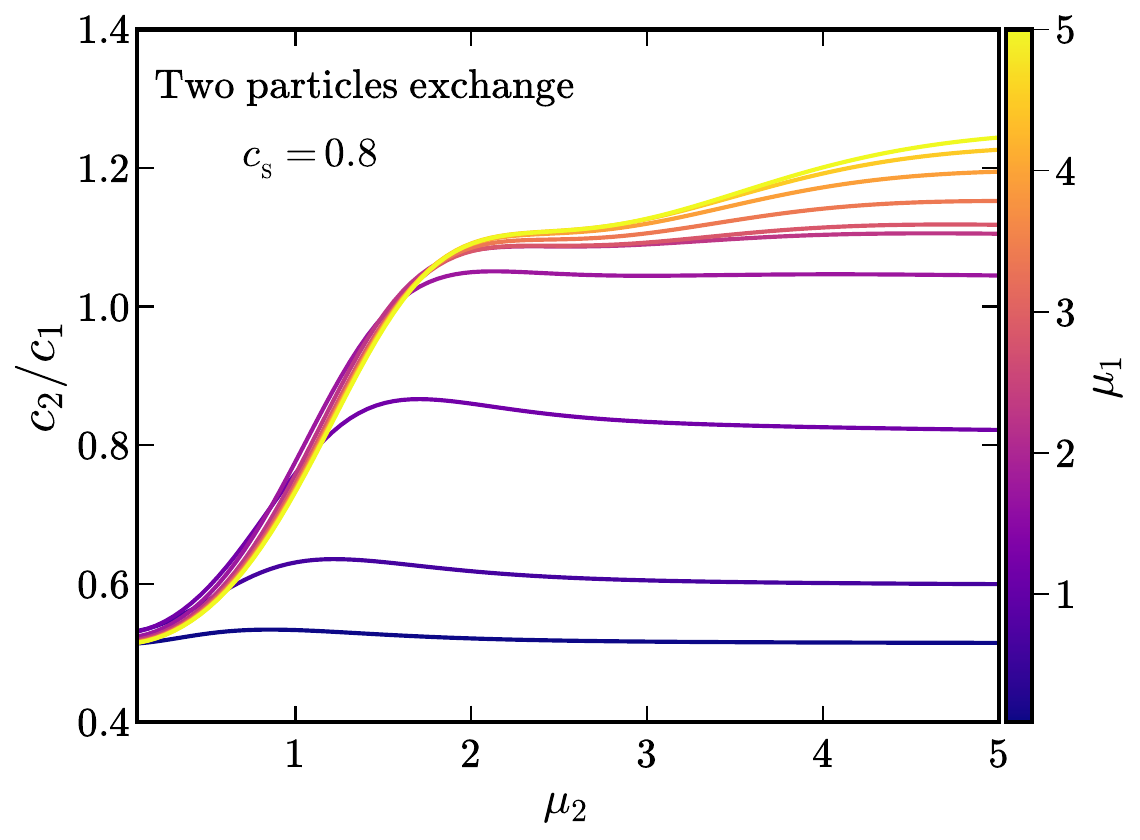}
    \includegraphics[width=0.49\linewidth]{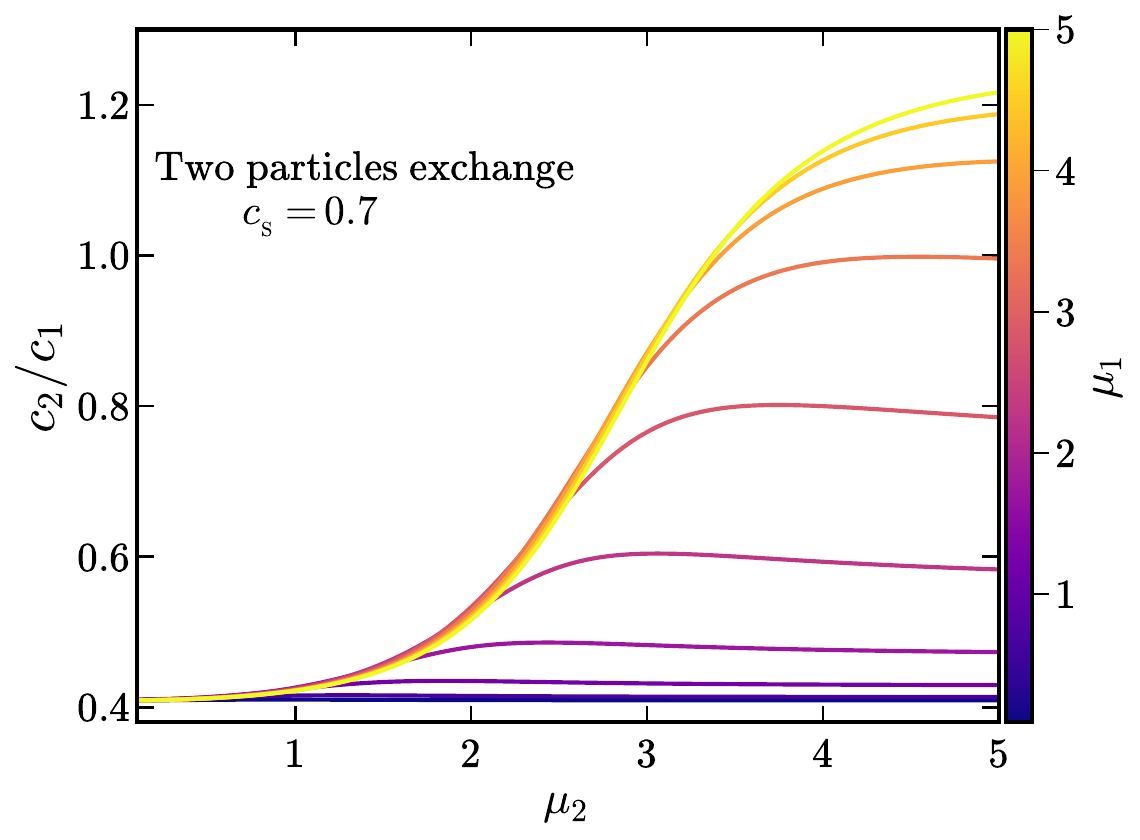}
    \caption{
Critical ratio $c_2/c_1$ separating regions with positive and negative equilateral bispectrum in the case of two-particle exchange. 
The panels correspond to different values of the sound speed: $\cs=1$ (top-left), $\cs=0.9$ (top-right), $\cs=0.8$ (bottom-left), and $\cs=0.7$ (bottom-right). 
In each panel, the curves correspond to fixed values of the mass parameter $\mu_1$ of the first exchanged particle, while the horizontal axis shows the mass parameter $\mu_2$ of the second particle. 
For positive values of $c_2/c_1$, the equilateral bispectrum is positive in the region above the curves, whereas for negative values of $c_2/c_1$ it is positive below the curves. 
The plots illustrate how the presence of multiple exchanged particles modifies the sign condition of the bispectrum and can significantly reduce the critical ratio required to obtain a positive equilateral configuration.
}
    \label{fig:twoexchange}
\end{figure*}
%%%%%%%%%%%%%%%%%%%%%%%%%%%%%%%%%%%%%%%%%%%%%%

%%%%%%%%%%%%%%%%%%%%%%%%%%%%%%%%%%%%%%%%%%%%%%
\begin{figure*}
    \centering
    \includegraphics[width=0.49\linewidth]{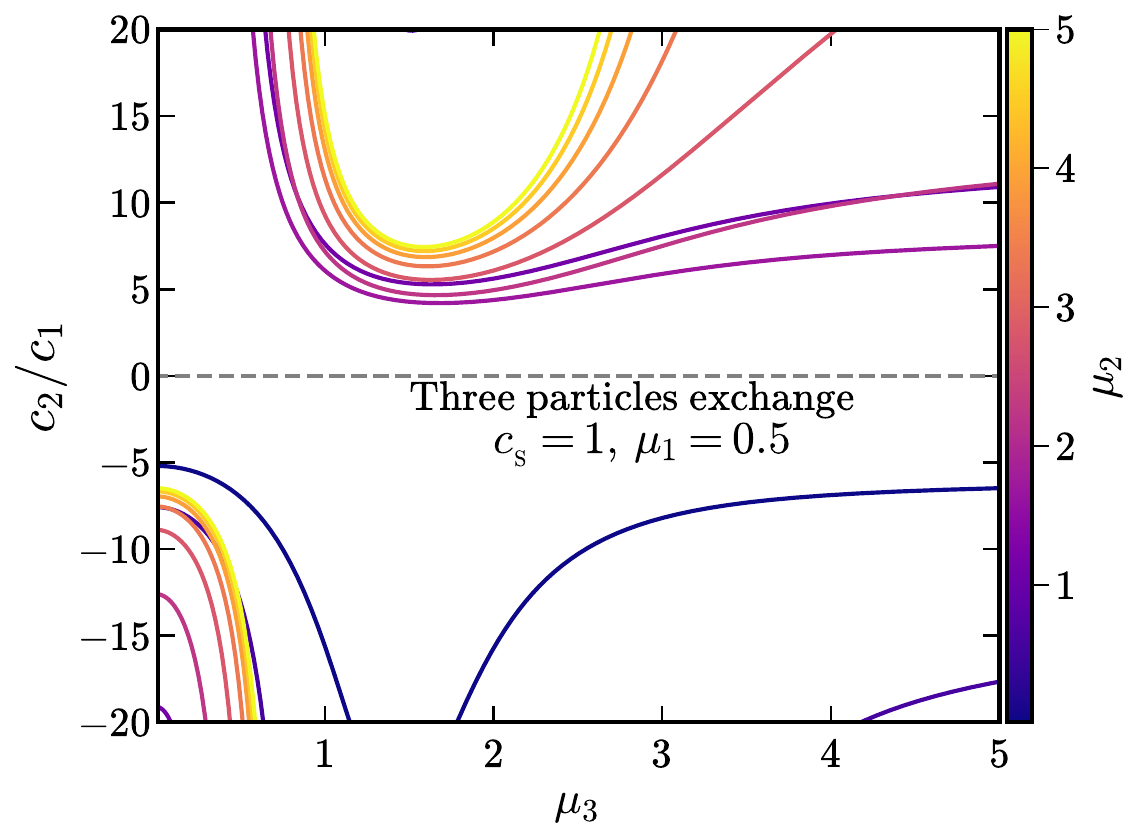}
    \includegraphics[width=0.49\linewidth]{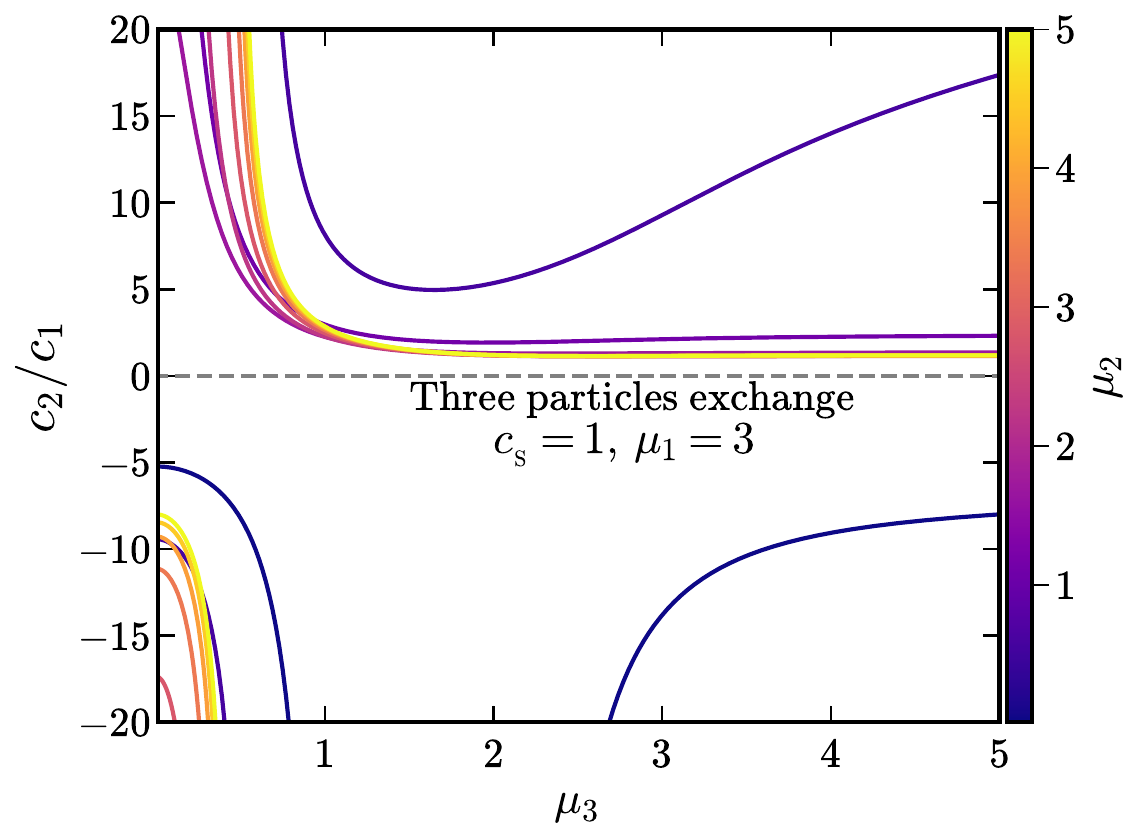}
    \includegraphics[width=0.49\linewidth]{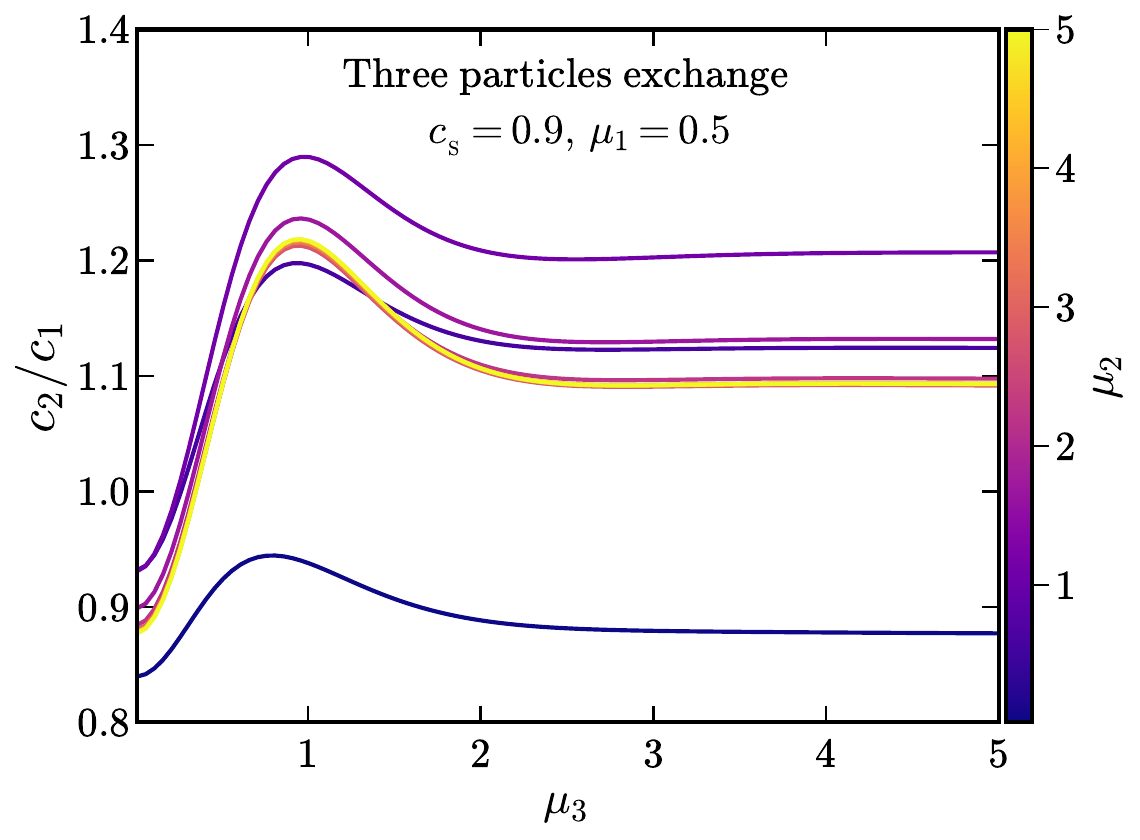}
    \includegraphics[width=0.49\linewidth]{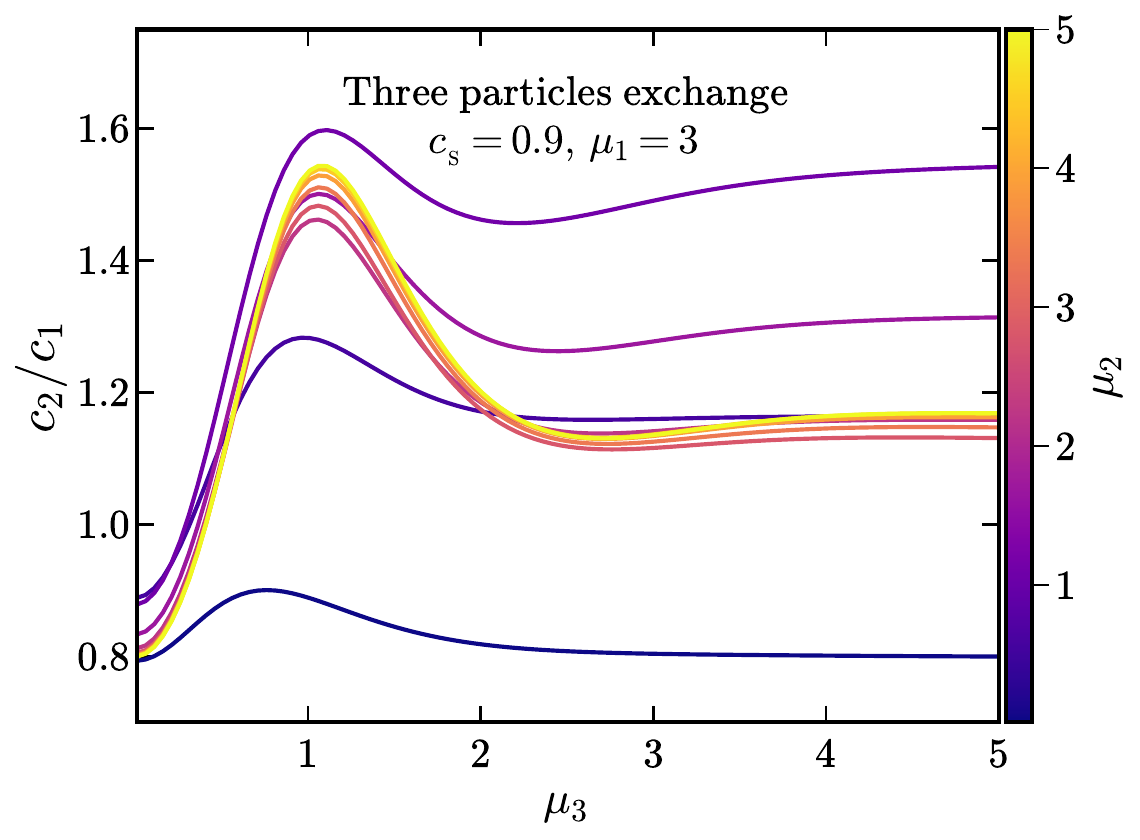}
    \includegraphics[width=0.49\linewidth]{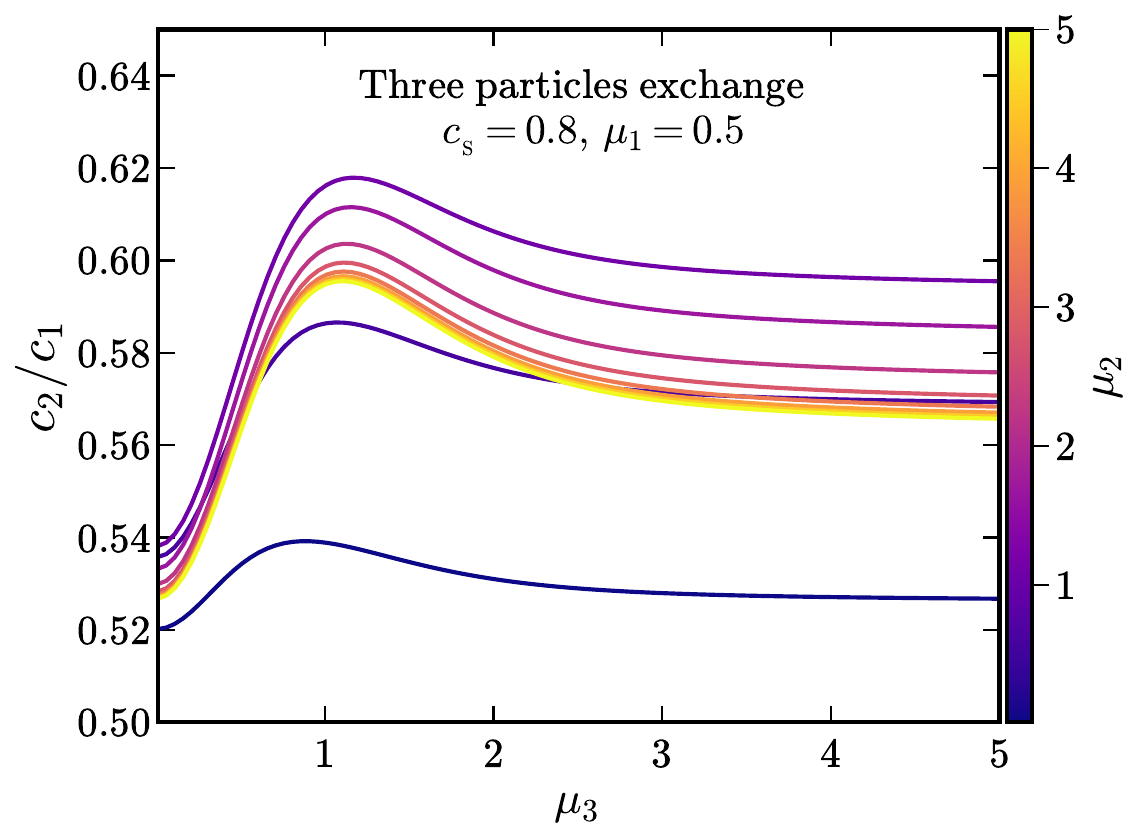}
    \includegraphics[width=0.49\linewidth]{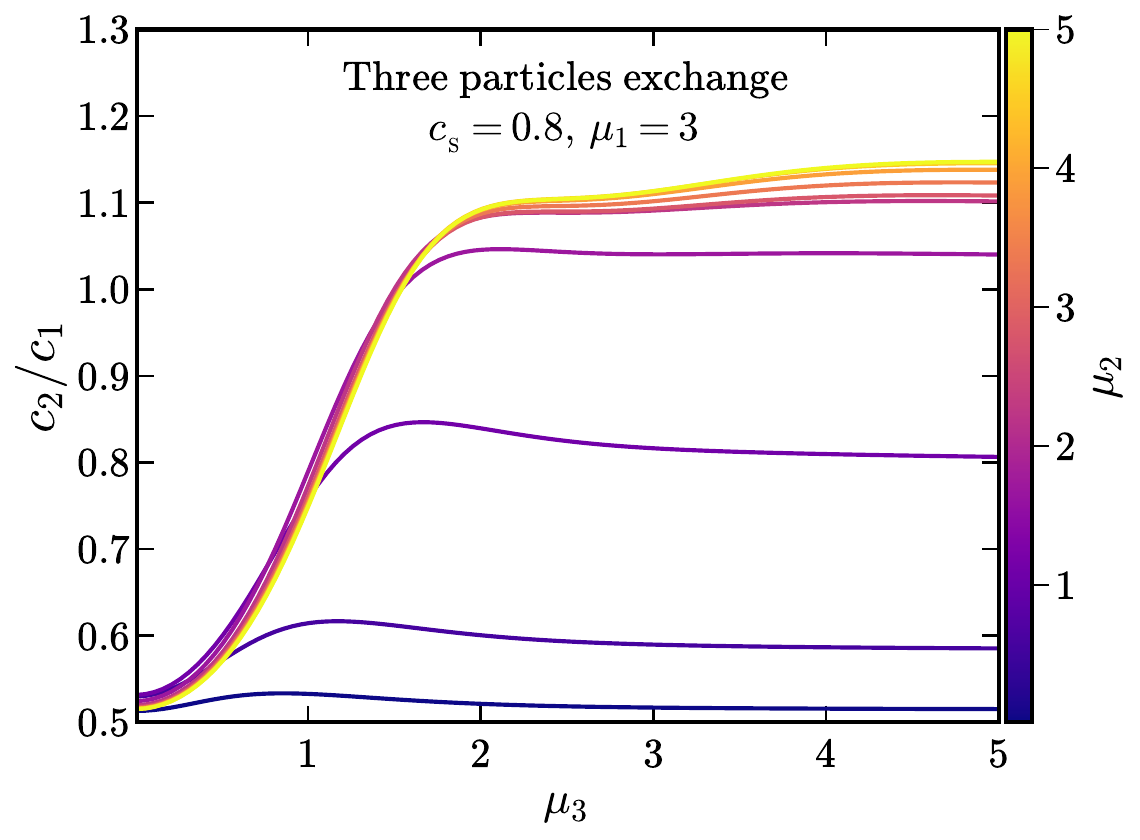}
    \caption{
Critical ratio $c_2/c_1$ separating regions with positive and negative equilateral bispectrum in the case of three-particle exchange. 
The rows correspond to different sound speeds: $\cs=1$ (top), $\cs=0.9$ (middle), and $\cs=0.8$ (bottom). 
The left and right columns correspond to two representative choices of the mass parameter of the first exchanged particle, $\mu_1=0.5$ and $\mu_1=3$, respectively. 
In each panel, the curves correspond to fixed values of $\mu_2$, while the horizontal axis shows the mass parameter $\mu_3$ of the third particle. 
For positive values of $c_2/c_1$, the equilateral bispectrum is positive in the region above the curves, whereas for negative values it is positive below the curves. 
The figure illustrates how the collective contributions of multiple exchanged particles can modify the sign structure of the bispectrum, allowing positive equilateral configurations even when the higher-order operator coefficient is smaller than the leading one.
}
    \label{fig:threeexchange}
\end{figure*}
%%%%%%%%%%%%%%%%%%%%%%%%%%%%%%%%%%%%%%%%%%%%%%

%%%%%%%%%%%%%%%%%%%%%%%%%%%%%%%%%%%%%%%%%%%%%%%%%%%%%%%%%%%%%%%%%%%%%%%%
\subsection*{Equilateral configuration: restricted vs enlarged operator basis}
%%%%%%%%%%%%%%%%%%%%%%%%%%%%%%%%%%%%%%%%%%%%%%%%%%%%%%%%%%%%%%%%%%%%%%%%

We now turn to the equilateral limit, where the sign structure of the bispectrum becomes particularly transparent.

\paragraph{Restricted operator basis ($c_2=0$):}

When the operator $\left(\delta g^{00}\right)^2\sigma$ is absent in the Lagrangian, the coefficients satisfy
\begin{align}
    \alpha_1 = \alpha_2 = c_1^2 .
\end{align}
In such a case, the overall sign of $c_1$ cancels from the bispectrum, and the equilateral configuration is determined entirely by the kinematical structure of $\hat F$ and its derivatives.
It has been found that for all real values of $\mu$ in the principal series~\cite{deRham:2025mjh},
\begin{align}
    B_\zeta^{\rm equil} < 0 .
\end{align}
The negativity of the equilateral bispectrum is therefore universal when the cubic structure is restricted to arise from the linear operator alone. 
In this regime, the sign is controlled purely by the de Sitter-invariant seed correlator and its kinematics.

\paragraph{Enlarged operator basis ($c_2\neq 0$):}

Once the $(g^{00}+1)^2\sigma$ operator is included, the coefficients satisfy
\begin{align}
    \alpha_1 \neq \alpha_2 ,
\end{align}
and the two kinematical structures enter with independent weights.
The left panel of Fig.~\ref{fig:ratio-massive} shows the purely kinematical parts of the two contributions (modulo the coupling factors).
We observe that:

\begin{itemize}
    \item The contribution associated with $(\partial_i\pi)^2\sigma$ remains negative for all $\mu$.
    \item The contribution associated with $\dot\pi^2\sigma$ can become positive for sufficiently small $\mu$.
\end{itemize}
The total equilateral bispectrum is therefore governed by the competition between these two structures.
Using the explicit expressions for $\alpha_1$ and $\alpha_2$, we obtain a critical ratio
\begin{align}
    \left(\frac{c_2}{c_1}\right)_{\rm crit}(\mu)
\end{align}
that separates regions of positive and negative equilateral bispectrum.
This boundary is shown in the right panel of Fig.~\ref{fig:ratio-massive}.

Thus, unlike the restricted case, the sign of the equilateral bispectrum is no longer universal.
It becomes sensitive to the relative weight of independent cubic structures allowed by the EFT.
Setting $c_2=0$ restores the universal negative sign. We further see that to get an equilateral bispectrum with a positive sign, $c_2$ does not have to be much larger. Even a ratio like $c_2/c_1\sim 1.5$ can turn the sign to be positive for large $\mu$.

% \color{blue}
%%%%%%%%%%%%%%%%%%%%%%%%%%%%%%%%%%%%%%%%%%%%%%%%%%%%%%%%%%%%%%%%%%%%%%%%
\subsection*{The case with $\cs<1$}
%%%%%%%%%%%%%%%%%%%%%%%%%%%%%%%%%%%%%%%%%%%%%%%%%%%%%%%%%%%%%%%%%%%%%%%%

So far, the bound on the ratio $c_2/c_1$ was derived for the case $\cs=1$. 
In that limit we found that a positive equilateral bispectrum can arise once the ratio $c_2/c_1$ crosses a critical value, with the sign change occurring around $\mu\sim0.6$. 
We now examine how this condition is modified when the sound speed differs from unity.

The sound speed affects the relative importance of time- and spatial-derivative interactions of the Goldstone mode. 
As a result, the contributions of the two cubic structures $\dot\pi^2\sigma$ and $(\partial_i\pi)^2\sigma$ to the bispectrum are altered when $\cs<1$, which in turn modifies the sign condition in the equilateral configuration.

The kinematical contributions of the time- and spatial-derivative operators for $\cs=0.8$ are shown in the left panel of Fig.~\ref{fig:csl1}. 
As in the $\cs=1$ case, the individual contributions from $\dot\pi^2\sigma$, $(\partial_i\pi)^2\sigma$, and the total combination $(\partial_\mu\pi)^2\sigma$ are plotted in blue dashed, red dashed, and black lines respectively. 
In contrast to the $\cs=1$ case, we find that for $\cs<1$ both $\dot\pi^2\sigma$ and $(\partial_i\pi)^2\sigma$ contributions are negative for all values of $\mu$, and consequently the total contribution for $c_2=0$ is also negative.

The right panel of Fig.~\ref{fig:csl1} shows the resulting critical ratio $c_2/c_1$ that separates regions with positive and negative equilateral bispectrum. 
We plot three representative values of the sound speed, $\cs=0.7$, $0.8$, and $0.9$, shown in yellow, red, and blue respectively. 
The shaded regions correspond to the values of $c_2/c_1$ for which the equilateral bispectrum becomes positive.

We observe that the critical ratio is particularly sensitive to the value of $\cs$ for small $\mu$, where the different curves are well separated. 
For larger values of $\mu$, however, the bounds for different $\cs$ gradually approach each other, indicating that the dependence on the sound speed becomes less pronounced in this regime.

%%%%%%%%%%%%%%%%%%%%%%%%%%%%%%%%%%%%%%%%%%%%%%%%%%%%%%%%%%%%%%%%%%%%%%%%

\subsection*{Exchange of multiple particles}

So far, we have studied the behaviour of the bispectrum in the equilateral limit arising from the exchange of a single massive scalar. 
In that case we derived the critical value of the ratio $c_2/c_1$ that determines whether the equilateral bispectrum is positive or negative, and examined how this bound depends on the sound speed $\cs$, particularly for small values of the mass parameter $\mu$. 

We now consider a more general situation in which multiple massive hidden-sector particles contribute to the exchange process~\cite{Xianyu:2023ytd,Aoki:2024uyi}. 
Such scenarios naturally arise in ultraviolet completions of inflation, where additional sectors may contain several heavy states—for example Kaluza–Klein towers, moduli fields in string compactifications, or multi-field sectors coupled to the inflaton. 
In these cases the observable bispectrum receives contributions from each exchanged particle.

If several scalars $\sigma_a$ with masses $\mu_a$ couple to the inflaton through the same EFT operators, the total bispectrum can be written schematically as
\begin{equation}
B_\zeta = \sum_a B_\zeta^{(a)}(\mu_a),
\end{equation}
where $B_\zeta^{(a)}$ denotes the contribution from the exchange of the $a$-th particle. 
Throughout this section we assume that the interactions arise from the same inflaton operators $(g^{00}+1)\sigma_a$ and $(g^{00}+1)^2\sigma_a$, so that the coefficients $c_1$ and $c_2$ are universal across the hidden-sector states. 
Under this assumption the equilateral bispectrum can be written schematically as
\begin{equation}
B^{\rm equil}_\zeta
=
\alpha_1 \sum_a \mathcal{I}_1(\mu_a,\cs)
+
\alpha_2 \sum_a \mathcal{I}_2(\mu_a,\cs),
\label{eq:multiexchange-schematic}
\end{equation}
where $\mathcal{I}_1$ and $\mathcal{I}_2$ denote the kinematical functions associated with the two cubic structures $\dot{\pi}^2\sigma$ and $(\partial_i\pi)^2\sigma$, respectively. 
Using $\alpha_1=c_1(c_1-4c_2)$ and $\alpha_2=c_1^2$, the condition $B^{\rm equil}_\zeta=0$ determines the critical ratio
\begin{equation}
\left(\frac{c_2}{c_1}\right)_{\rm crit}
=
\frac{1}{4}
\left[
1+\frac{\sum_a \mathcal{I}_2(\mu_a,\cs)}
{\sum_a \mathcal{I}_1(\mu_a,\cs)}
\right],
\quad
\sum_a \mathcal{I}_1(\mu_a,\cs)\neq 0 .
\label{eq:critical-ratio-multiparticle}
\end{equation}
This expression shows that the sign of the equilateral bispectrum is determined by the collective contribution of all exchanged particles to the two independent cubic structures. 
In particular, partial cancellations among the individual exchange contributions can significantly modify the sign condition. 
In the degenerate limit $\mu_a=\mu$, the critical ratio reduces to the single-particle result.

Let us first consider the case in which two particles are exchanged. 
Figure~\ref{fig:twoexchange} shows the resulting critical ratio of $c_2/c_1$ for different values of the sound speed $\cs$. 
The cases $\cs = 1$, $0.9$, $0.8$, and $0.7$ are displayed in the top-left, top-right, bottom-left, and bottom-right panels respectively. 
In each panel, every curve corresponds to a fixed value of $\mu_1$, while the ratio $c_2/c_1$ is plotted as a function of $\mu_2$. 
For positive values of $c_2/c_1$, the bispectrum is positive in the region above the curves, whereas for negative values it is positive below the curves. 
For $\cs=1$, certain combinations of $\mu_1$ and $\mu_2$ require $c_2$ to be significantly larger than $c_1$ to obtain a positive bispectrum. 
However, once $\cs<1$, the critical ratio typically remains $\cO(1)$ over a broad region of parameter space.

We next consider the exchange of three particles. 
The corresponding critical ratio is shown in Fig.~\ref{fig:threeexchange}. 
Here we illustrate representative cases with three values of $\cs$ and two choices of $\mu_1$. 
Each curve corresponds to a fixed value of $\mu_2$, while $c_2/c_1$ is plotted as a function of $\mu_3$. 
As in the two-particle case, the regions above or below the curves determine where the equilateral bispectrum becomes positive depending on the sign of $c_2/c_1$. 
For $\cs=0.9$ and $0.8$, the critical ratio again remains $\cO(1)$ across a wide range of masses.

An important qualitative feature of the multi-particle case is that the exchange contributions can partially cancel in the combination determining the sign of the bispectrum. 
Consequently, for many combinations of masses $\mu_a$, a positive equilateral bispectrum can arise even when the higher-order operator is subdominant, $c_2<c_1$. 
This effect is not present in the single-particle case and illustrates how the presence of multiple hidden-sector states can significantly modify the sign structure of the bispectrum.

Finally, it is worth noting that the contribution of heavier particles is exponentially suppressed for large $\mu$, reflecting the  Boltzmann suppression of heavy states in de Sitter space. 
As a result, the lightest exchanged particles typically dominate the sign condition, while heavier states mainly provide subleading corrections.

Overall, these results show that the presence of multiple hidden-sector particles enlarges the parameter space in which a positive equilateral bispectrum can arise. 
This further reinforces the conclusion that the sign of the bispectrum is sensitive not only to the operator structure of the EFT of inflation but also to the spectrum of heavy particles coupled to the inflaton.

\section{Conclusion}\label{sec:conclusion}

In this work, we revisited the inflationary bispectrum generated by the tree-level exchange of a massive hidden-sector scalar. 
Previous studies showed that when the interaction between the inflaton and the hidden sector arises solely from the leading EFT operator $(g^{00}+1)\sigma$, the equilateral bispectrum for principal-series scalar exchange is universally negative, independent of the sign of the coupling. 
This behavior reflects a kinematical universality inherited from the de Sitter-invariant seed correlator.

We showed that this universality relies crucially on the structure of the operator basis in the EFT of inflation. 
When only the leading coupling is present, the cubic interactions for the Goldstone mode appear with fixed relative weights and the equilateral bispectrum remains negative for all real values of the mass parameter $\mu$. 
However, once higher-order operators such as $(g^{00}+1)^2\sigma$ are included, the bispectrum receives contributions from independent cubic structures with different momentum dependence, notably $\dot\pi^2\sigma$ and $(\partial_i\pi)^2\sigma$. 
These contributions can compete in the equilateral configuration, leading to a loss of sign universality. 
We derived a critical ratio of interaction coefficients $c_2/c_1$ that separates regions of positive and negative equilateral bispectrum.

The characteristic oscillatory signal in the squeezed limit, $\sim (k_L/k_S)^{\pm i\mu}$, remains unchanged, continuing to encode the presence of the massive particle. 
However, the overall sign of the equilateral configuration becomes sensitive to the relative weight of independent cubic structures allowed by the EFT.

We further showed that the critical ratio depends on the sound speed $\cs$ and on the number of exchanged particles. 
For $c_s<1$, the bound on $c_2/c_1$ is modified, particularly for small $\mu$. 
In scenarios involving the exchange of multiple particles, a positive equilateral bispectrum can arise even when the higher-order operator coefficient satisfies $c_2<c_1$.

%Our results therefore demonstrate that the negativity of the equilateral bispectrum from massive exchange is not a generic prediction of inflationary collider physics, but rather a consequence of a restricted operator structure. 
%In more general EFTs of inflation, the sign of the equilateral configuration provides information about the relative importance of higher-order operators and the pattern of symmetry breaking in the effective action.

Several directions for future work remain. 
It would be interesting to extend this analysis to hidden-sector particles with spin, where angular dependence introduces additional structures. 
It would also be valuable to investigate whether positivity or analyticity constraints can further restrict the allowed parameter space of symmetry-breaking operators. 
Finally, exploring the observational consequences of a potential sign flip in the equilateral configuration may provide a useful probe of heavy physics during inflation.

%%%%%%%%%%%%%%%%%%%%%%%%%%%%%%%%%%%%%%%%%%%%%%%%%%%%%%%%%%%%%%%%%%%%%%%%%%%%%%%

\section*{Acknowledgments}

DG acknowledges support from the Core Research Grant CRG/2023/001448 of the Anusandhan National Research Foundation (ANRF) of the Gov. of India. FU acknowledges support from ANRF, National Post-Doctoral Fellowship (N-PDF).

%%%%%%%%%%%%%%%%%%%%%%%%%%%%%%%%%%%%%%%%%%%%%%%%%%%%%%%%%%%%%%%%%%%%%%%%%%%%%%%
%\clearpage
\balance

%\appendix
%\input{Sections/appendix}

\color{black}
\bibliographystyle{apsrev4-1}
\bibliography{Refe}
\end{document}